\documentclass[a4paper,twoside,10pt]{article}%
\usepackage[compress]{cite}
\usepackage{geometry}
\usepackage{authblk}
\usepackage{multirow}
\usepackage{hyperref}
\usepackage{url}
\usepackage{breakurl}
\usepackage{placeins}
\usepackage{amsmath}%
\newcommand\errors[2]{\genfrac{}{}{0pt}{}{#1}{#2}}
\usepackage{amsfonts}%
\usepackage{amssymb}%
\usepackage{graphicx}
\usepackage[ngerman,english]{babel}
\usepackage[T1]{fontenc}
\usepackage[utf8]{inputenc}
\usepackage{aeguill}

\usepackage{tabu}
\usepackage{booktabs}% for better rules in the table

\begin{document}
\title{Indications for Dzyaloshinskii-Moriya Interaction at the Pd/Fe Interface Studied by \textit{In Situ} Polarized Neutron Reflectometry}
\author[1,2]{Sina Mayr}
\author[1]{Jingfan Ye}
\author[2]{Jochen Stahn}
\author[3]{Birgit Knoblich}
\author[3]{Oliver Klein}
\author[4]{Dustin A. Gilbert}
\author[3]{Manfred Albrecht}
\author[1]{Amitesh Paul}
\author[1]{Peter B\"oni}
\author[1,5]{Wolfgang Kreuzpaintner}

\affil[1]{Technische Universit\"at M\"unchen, Physik-Department E21, James-Franck-Str.\ 1, 85748 Garching, Germany}
\affil[2]{Paul Scherrer Institut, CH-5232 Villigen PSI, Switzerland}
\affil[3]{Experimentalphysik IV, Institut f\"ur Physik, Universit\"at Augsburg, Universit\"atsstr.\ 1 Nord, 86159 Augsburg, Germany}
\affil[4]{Department of Materials Science and Engineering, University of Tennessee, Knoxville, TN 37996 USA}
\affil[5]{Guangdong Technion-Israel Institute of Technology, 241 Daxue Road, Jinping District, Shantou, Guangdong Province, China}
\date{April 17, 2019}

\maketitle

\begin{abstract}
Using \textit{in situ} polarized neutron reflectometry, the depth resolved evolution of the magnetism and structure in a Pd/Fe/Pd trilayer thin-film is measured during growth. The initial film structure of Pd/Fe shows a small proximity induced magnetism in the underlayer and a magnetization in the Fe layer of $\approx1.6$\,$\mu_{\text{B}}$ per Fe atom, less than the expected bulk value of $2.2$\,$\mu_{\text{B}}$. Deposition of the Pd capping layer initially follows an island-like growth mode with subsequent coalescence. With increasing Pd deposition the Fe moment and the proximity-induced magnetism in the Pd capping layer decrease. After final deposition of the Pd capping layer, the magnetic profile is structurally and magnetically symmetric across the Fe layer, with magnetism induced in Pd up to 0.92 \,nm from the interface. Throughout the Pd deposition the Pd/Fe/Pd trilayer structure is becoming increasingly symmetric, a fact which points to a Dzyaloshinskii-Moriya interaction as a likely cause of the observed magnetic behavior.
\end{abstract}

% insert suggested keywords - APS authors don't need to do this
%\keywords{}

%\maketitle must follow title, authors, abstract, and keywords
%\maketitle

% figures should be put into the text as floats.
% Use the graphics or graphicx packages (distributed with LaTeX2e)
% and the \includegraphics macro defined in those packages.
% See the LaTeX Graphics Companion by Michel Goosens, Sebastian Rahtz,
% and Frank Mittelbach for instance.

\section{Introduction}
Thin films and heterostructures exhibit a variety of fascinating electronic, magnetic, and optical properties \cite{doi:10.1038/nmat2804,doi:10.1038/nmat3870,doi:10.1126/science.aaa5198,doi:10.1126/science.1146006,doi:10.1038/nmat1931,doi:10.1038/nmat2049,doi:10.1038/nmat2012,doi:10.1038/nmat2916,doi:10.1103/PhysRevLett.108.267201} and are therefore indispensable for scientific and technological applications. Most of the functional characteristics of layered structures are determined by the processes taking place during their preparation by thin film deposition, viz.\ when the sample structure, material stoichiometry and defect population are defined. For a fundamental understanding of magnetism on an atomic scale, it is therefore crucial to investigate the evolution of magnetism in thin layers and heterostructures \textit{in situ} during growth and to correlate the magnetic properties with the corresponding microstructure. \textit{In situ} studies are particularly critical when the system possesses proximity effects such as induced magnetism.
 
The Fe/Pd thin film system is known to show strong proximity induced magnetism and has been widely studied experimentally \cite{doi:10.1103/PhysRevB.90.104403, doi:10.1103/PhysRevLett.65.1156, doi:10.1103/PhysRevB.44.2205, doi:10.1016/S0304-8853(96)00480-5, doi:10.1016/0304-8853(94)90652-1, doi:10.1103/PhysRevB.51.6364, doi:10.1103/PhysRevLett.72.2247, doi:10.1103/PhysRevB.65.014423} and theoretically \cite{doi:10.1103/PhysRevB.51.6364, 0953-8984-19-24-246213, doi:10.1103/PhysRevLett.72.2247, doi:10.1103/PhysRevB.65.014423, doi:10.1103/PhysRevB.52.12516, doi:10.1007/BF00616980}. In the Fe/Pd system, proximity induced magnetism has been reported in the Pd layer up to 2\,nm \cite{doi:10.1103/PhysRevB.90.104403} from the Fe interface, with an induced magnetization of $0.3 - 0.4$\,$\mu_{\text{B}}$/atom$^{\text{Pd}}$ at the interface \cite{doi:10.1016/S0304-8853(96)00480-5, doi:10.1007/BF00616980, doi:10.1103/PhysRevB.51.6364}. 

Further, at the interface between a heavy metal (HM) element with strong spin-orbit coupling and a ferromagnetic transition metal (FM), magnetic spin structures develop chiral domain walls, spirals or skyrmions due to the interfacial Dzyaloshinskii-Moriya interaction (IDMI) \cite{doi:10.1038/nmat4934,doi:10.1103/PhysRevLett.118.147201,doi:10.1103/PhysRevLett.120.157204,doi:10.1103/PhysRevMaterials.2.044404, doi:10.1002/adma.201800199,doi:10.1103/PhysRevB.97.024404,doi:10.1103/PhysRevLett.120.157204,doi:10.1103/PhysRevB.96.060410,doi:10.1038/s41598-017-17137-z}. The presence of IDMI is not related to the existence of proximity-induced magnetization \cite{doi.10.1103/PhysRevLett.115.267210}. Accordingly, in systems that show both IDMI and proximity-induced magnetism, it is not clear what the magnetic structure at the interface will be, especially when the surfaces are asymmetric. Furthermore, the HM layer can be used to generate strong spin-orbit torques, arising from the Spin-Hall-effect \cite{doi:10.1038/nature19820}, allowing the manipulation of these interfacial magnetic structures without affecting the induced magnetization away from the interface. This makes such systems highly promising candidates to realize high-speed and energy efficient memory devices and offer tremendous opportunities for research and technological applications.  

In this work, the evolution of magnetism in a polycrystalline Pd(11\,nm)/Fe(0.41\,nm)/Pd(72\,nm) trilayer heterostructure grown on a Si substrate is investigated. Films were grown by conventional direct current (dc) magnetron sputtering and investigated \textit{in situ} by polarized neutron reflectometry (PNR). The step-wise deposition of Pd onto a Pd/Fe bilayer system initially occurs by island growth before becoming a continuous film. The induced magnetism in both the Pd capping layer and Pd underlayer are separately resolved and quantified as Pd is deposited. Surprisingly, the magnetism in the Pd films is initially asymmetric, with the thin capping layer having an induced moment of 0.6\,$\mu_{\text{B}}$ per atom, while the induced magnetism in the underlayer is 0.2\,$\mu_{\text{B}}$ per atom ($\sim 65\%$ smaller). Increasing the Pd capping layer thickness motivates a symmetric structure, and results in a magnetic symmetry. The interfacial Dzyaloshinskii-Moriya interaction is implicated as a likely origin of the observed behaviour.

\section{Experimental Details}
Films of Pd(11\,nm)/Fe(0.41\,nm)/Pd(72\,nm) were grown by dc magnetron sputtering on a $2 \times 2$\,cm$^2$ Si(001) substrate in an ultrahigh-vacuum chamber (p$_{base}=5 \times 10^{-9}$\,mbar) installed at the AMOR reflectometer at Paul Scherrer Institute, Villigen \cite{doi:10.1016/j.nima.2016.03.007}. The deposition system is equipped with three 2\,$^{\prime\prime}$ sputter sources which can be positioned above the sample surface without breaking the vacuum. This allows the Pd and Fe to be deposited without exposing the sample to atmosphere or realigning it in the neutron beam. Control over the layer thicknesses is achieved via the opening times of a deposition shutter.

Sample growth was performed at room temperature without any heating or cooling of the substrate. Pd was sputtered from a 99.99\% pure target using an ultra-high purity (7N) Ar working gas at a pressure of $3.37 \times 10^{-3}$\,mbar and with 20\,W of dc sputtering power, resulting in a deposition rate of $0.33$\,nm/s. Fe was deposited at an Ar pressure of $4.69 \times 10^{-3}$\,mbar from a 99.95\% pure sputtering target at 20\,W, resulting in a deposition rate of $0.13$\,nm/s. The Pd underlayer and Fe layer were grown in single steps while the 11\,nm Pd capping layer was grown in 18 steps with approximately $0.6$\,$\mu$g/cm$^2$ of Pd deposited per step.

The evolution of the depth-averaged nuclear density and magnetization was followed \textit{in situ}, with PNR measurements being performed after each of the deposition steps. PNR is a grazing-incidence neutron scattering technique with high spatial magnetic and nuclear sensitivity \cite{doi:10.1051/sfn/20141302004,doi:10.1088/0953-8984/15/5/306}. 

Except for very few demonstration cases, PNR measurements are all performed on films after growth (i.e.\ \textit{ex situ}), and thus emergent behavior during thin-film growth could not be investigated by this technique. Facilitated by recent developments \cite{doi:10.1016/j.nima.2017.11.086,doi:10.1103/PhysRevApplied.7.054004}, PNR can now also be applied as an \textit{in situ} technique (\textit{i}PNR). By growing samples at the neutron beamline, \textit{i}PNR allows the evolution of the structural and magnetic properties of the entire film to be captured as a function of the layer thicknesses, one deposition step after the other. For clarity, \textit{i}PNR captures the magnetic and nuclear depth profile of the entire film at each measurement, with spatial sensitivity along the sample's thickness. Therefore, the evolution of the depth profile as a function of deposited Pd thickness is resolved. The major advantage of \textit{i}PNR for the study presented here lies with the simultaneous accessibility to the magnetic properties in Pd on both sides of the Fe layer.

To perform these measurements in a reasonable timeframe, prototype focusing Selene neutron optics \cite{doi:10.1016/j.nima.2010.06.221, doi:10.1051/epjap/2012110295, doi:10.1016/j.nima.2016.03.007} were used, with a neutron wavelength band of 4 -- 10\,\AA\ and a neutron beam divergence of 1.6\,$^{\circ}$. With these settings the resolution increases from $\frac{\Delta q}{q} \approx 4.5\%$ in the regime of total reflection to a quasi-static value of $\frac{\Delta q}{q} \approx 2.3\%$ for $q_z \gtrsim 0.2$\,nm$^{-1}$. Beam polarization was realized by the transmittance of the neutrons through a $m = 4.2$ Fe/Si multilayer polarizer with a logarithmic spiral shape \cite{doi:10.1088/1742-6596/862/1/01200}. The neutron polarization was selected by an RF spin-flipper. In-vacuum guide fields perpendicular to the scattering plane maintained the neutron polarization up to the sample position. A magnetic field of 70\,mT was applied in-plane to saturate the sample using permanent magnets. The \textit{i}PNR data acquisition times were approximately 35\,min for each spin direction, which at the given vacuum quality was sufficient to rule out any contamination from the residual gas. 

A theoretical model was fitted to the experimental \textit{i}PNR data using the SimulReflec software \cite{SimulReflec}. The errors of the fit parameters are estimated by a 5\% increase over the optimum figure of merit $FOM \approx\sum{\left|\ln{R_{\text{fit}}}-\ln{R_{\text{meas}}}\right|}$ on independent variation of a single parameter \cite{doi:10.1107/S0021889807045086}, where $R_{\text{fit}}$ is the fitted and $R_{\text{meas}}$ the measured reflectivity, respectively. 

The theoretical models include a natural SiO$_2$ layer on the surface of the Si substrate on which the Pd(11\,nm)/Fe(0.41\,nm)/Pd(72\,nm) trilayer structure is grown. These fit parameters are given in Table \ref{tab:subst_fit_param}.

\begin{center}	
	\begin{table}[h]
		\caption{PNR fit parameters for the Si substrate, the Pd seed, and the Fe layer. For the number density ($n$), also the ratio $n/n_{\text{bulk}}$ is given.}
		\resizebox{\textwidth}{!}{%
				\begin{tabular}{| l || c | c | c | c | }
				\hline
											& thickness ($d$)																																& number density ($n$) 																																								& $n/n_{\text{bulk}}$  						&	rms roughness ($\sigma$)\\ 
				\hline
				\hline
				Si  					& 0.65\,mm 																																			& $n^{\text{Si}}=4.81 \left(\errors{+0.37}{-0.39} \right) \times 10^{22}$\,cm$^{-3}$ 									& 0.97 														& $				\sigma^{\text{Si}}=0.26 \left(\errors{+0.14}{-0.16} \right)$\,nm \\ 
				\hline
				SiO$_2$ layer & $d^{\text{SiO}_2}=0.99 \left(\errors{+0.51}{-0.56}\right) $\,nm 							& $n^{\text{SiO}_2}=2.58 \left(\errors{+0.41}{-0.26}\right)\times 10^{22}$\,cm$^{-3}$ 								& 0.96 														& $\sigma^{\text{SiO}_2}=0.58 \left(\errors{+0.24}{-0.19} \right)$\,nm \\
				\hline
				\hline
				Pd seed 			& $d^{\text{Pd}_{\text{seed}}}=71.93 \left(\errors{+1.09}{-0.79} \right) $\,nm 	& $n^{\text{Pd}_{\text{seed}}} = 5.66 \left(\errors{+0.16}{-0.05} \right) \times 10^{22}$\,cm$^{-3}$ 	& 0.83 														& $\sigma^{\text{Pd}_{\text{seed}}}=1.76 \left(\errors{+0.06}{-0.10} \right)$\,nm\\
				\hline
				Fe layer 			& $d^{\text{Fe}}= 0.41 \left(\errors{+0.11}{-0.12} \right)$\,nm 								& $n^{\text{Fe}} = 7.97 \left(\errors{+0.31}{-0.29} \right) \times 10^{22}$\,cm$^{-3}$ 								& 0.94 														& $\sigma^{\text{Fe}} = 1.73\ (\pm 0.06)$\,nm \\
				\hline
			\end{tabular}
		}
		\label{tab:subst_fit_param}
	\end{table}
\end{center}

The Fe layer was simulated such that its structural and magnetic thickness was identical. For Pd, the best agreement with the experimental data was obtained by allowing a region of up to 0.92\,nm from the Fe interface to carry a magnetic moment. This magnetic regime agrees well with previous reports on induced magnetism in the first 2 \cite{doi:10.1103/PhysRevB.51.6364} to 4 monolayers \cite{doi:10.1016/S0304-8853(96)00480-5}. For the fitting process the 0.92\,nm regions on either side of the Fe layer were split into four equally thick regimes of 0.23\,nm, each carrying its own magnetization. These magnetic regimes are identified as Pd${-4}$, Pd${-3}$, Pd${-2}$, Pd${-1}$ (below the Fe layer) and Pd${+1}$, Pd${+2}$, Pd${+3}$ and Pd${+4}$ (covering the Fe) with Pd${-1}$ and Pd${+1}$ being at the interface to the Fe and Pd${-4}$ and Pd${+4}$ being the interface to the assumed non-magnetic part of the Pd layers. The theoretical model is shown in the inset of Figure \ref{fig:Kreuzpaintner_Figure_1}. It is noted that attempts to simulate the PNR data on the basis of a different model resulted in a less perfect agreement to the \textit{i}PNR data (viz.\ models based on no magnetization in the Pd \cite{doi:10.1103/PhysRevB.51.6364} or larger regimes of induced magnetization), or in non-physical values of the Fermi-vectors for models based on the assumption of RKKY-like oscillatory magnetic behavior in the Pd.

Transmission electron microscopy (TEM) measurements were performed \textit{ex situ} on the fully grown sample using a field-emission JEOL\,2100\,F microscope at 200\,kV, equipped with an ultrahigh resolution pole-piece. The point-to-point resolution specified by the manufacturer is 0.194\,nm. Note that a value for the interfacial rms roughness is obtained from the decay of the \textit{i}PNR reflectivity curves. It statistically describes the deviation from the mean interface level over the illuminated sample area of $20 \times 2$\,mm$^2$, but does not allow the lateral in-plane length scales at which an interface ``appears'' rough, to be assessed. For this, off-specular \textit{i}PNR data would be required, which is not obtainable if the Selene neutron optics is applied. Consequently, the TEM results are used to locally identify interface and layer quality, and the microstructure of the fully grown sample.

\section{Results and Discussion}
\begin{figure}
\centering
\includegraphics[width=0.5\textwidth]{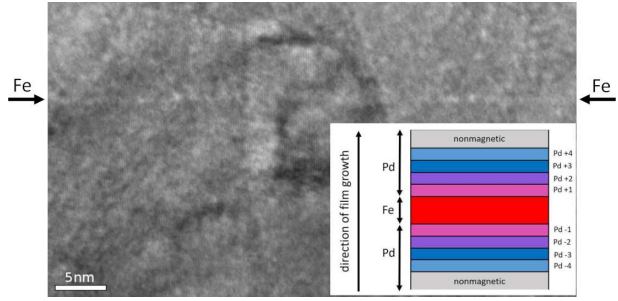}
	\caption{\label{fig:Kreuzpaintner_Figure_1} TEM image showing the Fe layer buried in Pd. The sharp interfaces between Fe and Pd are visible. The inset shows the model for the magnetic structure of the sample used for the fitting process with the Fe layer (red) and the adjacent Pd sublayers Pd-1 to Pd-4 and Pd+1 to Pd+4.}
\end{figure}

Images obtained by transmission electron microscopy, Figure \ref{fig:Kreuzpaintner_Figure_1}, show that the Pd and Fe layers are polycrystalline. The mass contrast of the Fe and Pd allows the identification of the Fe(0.41\,nm) layer as a horizontal white line in the figure. The sharp boundary of the Fe layer confirms the high-quality growth of the film with minimal interdiffusion.

\begin{figure}
\centering
\includegraphics[width=0.5\textwidth]{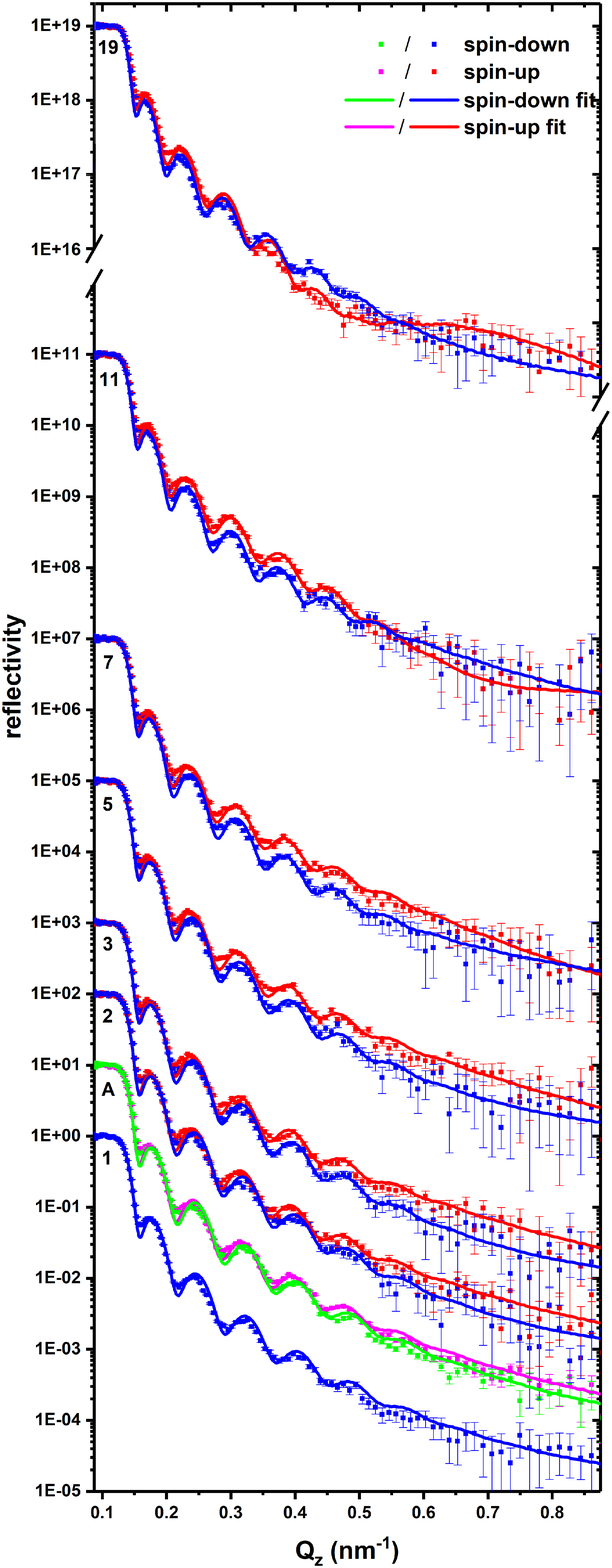}
	\caption{\label{fig:Kreuzpaintner_Figure_2} Measured reflectivity curves are shown in red and blue for spin up and spin down, respectively, together with their best fits as solid lines. The numbers denote the Pd deposition steps $i$. The letter $A$ denotes the deposition step $i=A$ in which Fe was deposited. For this deposition step, the spin up and spin down reflectivities are shown in magenta and green, respectively. $i=1$ corresponds to the deposition of the Pd underlayer. It was followed by the deposition step with Fe ($i=A$). Steps $2 \leq i \leq 19$ show the reflectivity profiles after deposition of Pd layers on top of the Fe layer.}
\end{figure}

\begin{figure}
\centering
\includegraphics[width=0.5\textwidth]{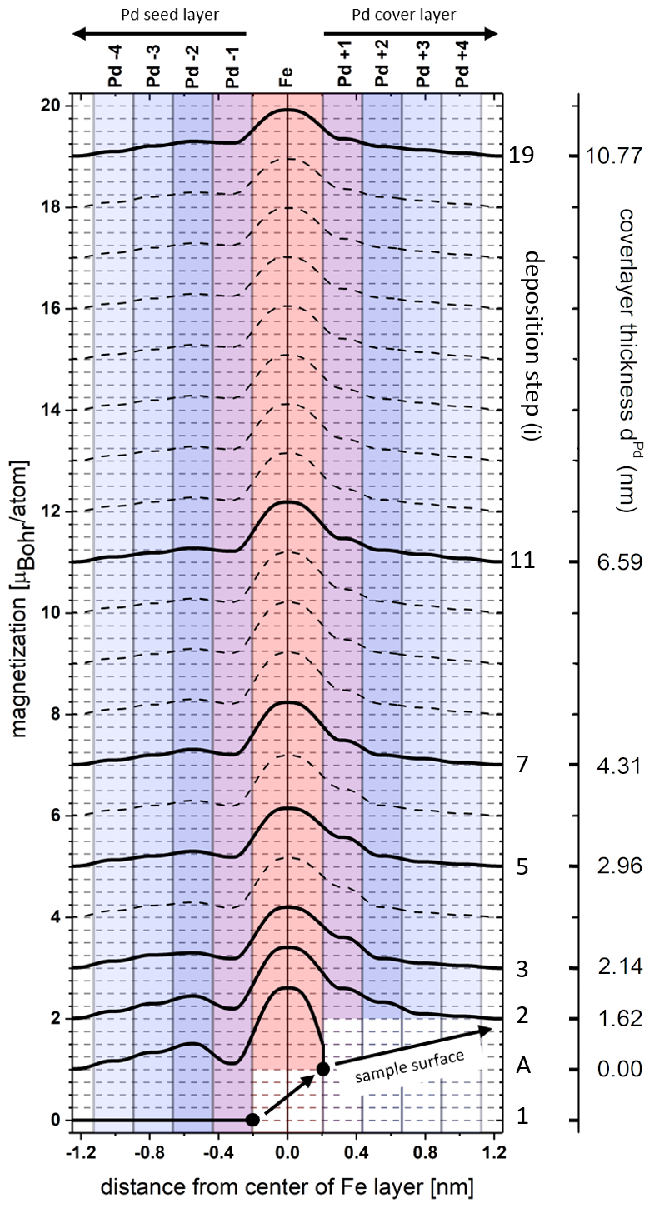}
	\caption{	\label{fig:Kreuzpaintner_Figure_3} Magnetization profiles as a function of deposition step and distance from the center of the Fe layer, obtained from the best fit of the profiles of the magnetic scattering length density. Each curve is vertically shifted by 1\,$\mu_{\text{B}}$/atom for better visibility. As guide to the eye, the dashed lines show interpolated curves. The magnetization in Pd extends over distances of up to 0.92\,nm to either side of the Fe layer -- split up into four magnetic sublayers -- and a nonmagnetic rest. The assumed magnetic regions are referred to as Pd${-4}$, Pd${-3}$, Pd${-2}$, Pd${-1}$ (Pd seed layer) and Pd${+1}$, Pd${+2}$, Pd${+3}$ and Pd${+4}$ (Pd capping layer). In the initial stages of Pd capping layer growth a strong side asymmetry for the magnetization of Pd${-1}$ and Pd${+1}$ is observed. This asymmetry vanishes as the structural symmetry of the Pd/Fe/Pd trilayer increases with each deposition step. With increasing symmetry in the magnetization profile, the magnetization in the Fe layer is also slightly reduced.}
\end{figure}

The \textit{i}PNR data overlaid with best-fit theoretical reflectivity curves as a function of deposition step \textit{i} is shown in Figure \ref{fig:Kreuzpaintner_Figure_2}. The magnetic depth profiles which fit best the experimental reflectivity of each measured deposition step are shown in Figure \ref{fig:Kreuzpaintner_Figure_3}. Here, the magnetic moments are plotted in units of $\mu_{\text{B}}$.  
The evolution of the nuclear (e.\,g.\ structural) and magnetic depth profiles versus deposition step are discussed below. 

\paragraph{Structural Evolution of the Sample During Growth}

The initial Pd seed layer ($i=1$) shows a fitted structural thickness of $\sim 72 $\,nm and an interfacial rms roughness of $\sim 1.8 $\,nm. 
The Fe layer (step $i=A$) was found to have a thickness of $\sim 0.4 $\,nm and a number density of $ \sim 8 \times 10^{22}$\,cm$^{-3}$, which is between the number densities found for liquid ($\approx 7.5 \times 10^{22}$\,\,cm$^{-3}$) and single crystalline bulk Fe ($\approx 8.5 \times 10^{22}$\,cm$^{-3}$). In combination with the sharp and smooth appearance of the Fe layer in the TEM image, this indicates that the ultra thin Fe layer was grown as a continuous layer onto the Pd underlayer. The rms interface roughness of Fe also reproduces the interface roughness of the underlying Pd layer nearly unaltered. The values (Table \ref{tab:subst_fit_param}) for the Fe film and Pd underlayer agree well with the nominal structure and the expected values for polycrystalline Pd and Fe. 

\begin{figure}
\centering
\includegraphics[width=0.5\textwidth]{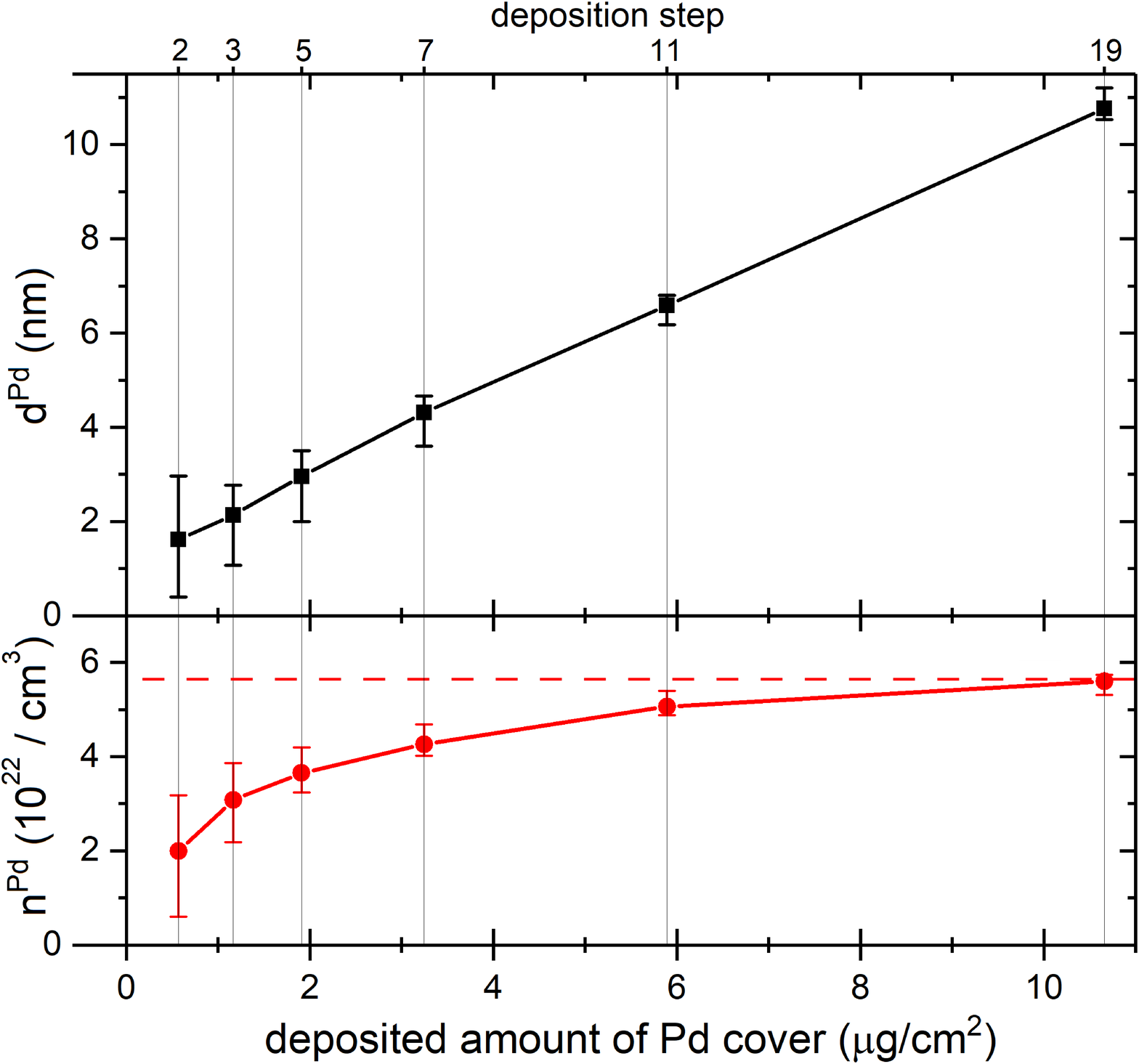}
	\caption{\label{fig:Kreuzpaintner_Figure_4} Evolution of the layer thickness ($d^{\text{Pd}}$) and number density ($n^{\text{Pd}}$) of the Pd capping layer as a function of deposition step $i$ and deposited amount of Pd. The connecting lines are a guide to the eye. The dashed line gives the number density $n^{\text{Pd}_{\text{seed}}} \approx 5.66 \times 10^{22}$\,cm$^{-3}$ of the underlying Pd seed layer.}
\end{figure}

The evolution of the fitting parameters for the Pd capping layer (Figure \ref{fig:Kreuzpaintner_Figure_4}) indicates a significantly different growth mode as compared to the Fe, which formed a continuous layer within the single deposition step. In particular the nuclear scattering length density of the Pd capping layer was initially much lower than bulk, and asymptotically approaches the bulk value with increasing thickness. This indicates that the Pd capping layer started its growth as islands, with a continuously progressing coalescence with each deposition step until a number density value of $\approx 5.7 \times 10^{22}$\,cm$^{-3}$ is reached. 
The simultaneous decrease of surface roughness relative to the thickness of the layers in the early growth steps also traces the coalescence of the initial Pd islands. The increasing coalescence, and with it the symmetry of the sample structure relative to the Fe layer influences the magnetization profile on both sides of the Fe layer.

\paragraph{Magnetism in the Fe layer}
The magnetization profile of the bi-layer structure, consisting of the uncapped Fe on Pd (viz.\ $d^{Pd}=0$\,nm, growth step $i = A$) was measured with a converged magnetic moment of M$^{\text{Fe}}_{i=A} \approx 1.6$\,$\mu_{\text{B}}$. With progressing deposition of the Pd capping layer onto the Fe layer, M$^{\text{Fe}}$ successively decreases to a final value of M$^{\text{Fe}}_{i=19} \approx0.9$\,$\mu_{\text{B}}$/atom$^{\text{Fe}}$ (Figure \ref{fig:Kreuzpaintner_Figure_3}). The lower magnetization value of the Fe layer as compared to the bulk value of $2.22$\,$\mu_{\text{B}}$/atom$^{\text{Fe}}$ for bcc-Fe may be caused by the Fe layer being too thin to exhibit a bulk-like magnetization. In particular, an enhanced magnetic moment for Fe as reported in \cite{doi:10.1103/PhysRevB.51.6364} could not be observed. Note that in \cite{doi:10.1103/PhysRevB.51.6364} PNR was carried out with the sample cooled to below 20\,K and that in the model used for the analysis of the PNR data, the Pd was not allowed to carry any induced magnetization. This may explain the discrepancy to the findings presented here. It is noted that a theoretical model with an enhanced magnetic moment within the Fe alone and without any induced magnetization in Pd cannot reproduce the \textit{i}PNR data shown in Figure \ref{fig:Kreuzpaintner_Figure_2}.

A key observation of \textit{i}PNR is that the measured saturation magnetization in Fe seems to decrease with increasing Pd thickness. In the presented measurement configuration, \textit{i}PNR is solely sensitive to the net in-plane projection of magnetic moments, and thus the decrease may be the result of a spin-reorientation to the out-of-plane direction, the formation of domains, or an authentic reduction of the Fe magnetization. Considering the first of these possibilities, an in-plane magnetic field of 70\,mT was applied during the measurements to saturate the magnetic moments into the film plane. The decrease in the measured Fe magnetization would therefore indicate an increase in the anisotropy away from the field, presumably into the out-of-plane direction. However, previous works have reported that the Pd/Fe system at room temperature has a dominant in-plane anisotropy for all thicknesses \cite{doi:10.1063/1.345954}, while other works performed at low-temperatures have reported that the out-of-plane anisotropy decreases with increasing thickness of the Pd capping layer \cite{doi:10.1103/PhysRevB.85.224406}. Based on these results, the measured projection of the magnetization should \textit{increase} with Pd thickness, contrary to our observations. 

Growing the Pd capping layer by room-temperature sputtering, rather than e.\,g.\ e-beam evaporation \cite{doi:10.1103/PhysRevB.85.224406}, results in initial island growth with increasing coalescence as more Pd is deposited. The increasing contact area between the Pd and Fe 
may change the magnetization. However, as discussed above, the increased interaction should promote an in-plane orientation in the Fe, and thus would also \textit{increase} the measured magnetization, contrary to the observed decrease. We therefore conclude that the observed trend is likely not caused by out-of-plane spin reorientation due to interfacial perpendicular anisotropy. 

We suggest that the increasing layer symmetry and increasing density of the Pd cap are the primary mechanisms which influence the magnetization in the Fe layer. One explanation is that an increased hybridization between the 3d Fe and 4d Pd valence electrons takes place as the Pd capping layer thickness increases. This hybridization will initially scale with the thickness due to the increased number of interfacial neighbors, and would result in a modification of the density of states at the Fermi level according to the Stoner model of magnetization. This hybridization may be sufficient to explain the decrease in magnetization of the Fe layer. Additionally, the increased hybridization will enhance the Dzyaloshinskii-Moriya interactions (DMI). As discussed below, the DMI promotes a curling of the magnetization away from a saturated configuration, which would appear as a reduction in the magnetization when probed with PNR. 

\paragraph{Magnetism in the Pd layers}
In contrast to most previous studies, \textit{i}PNR allows simultaneous spatially resolved access to the magnetic properties in both the top and bottom Pd layers, in addition to the Fe layer. The increased spatial resolution is demonstrated to be a great asset, as a strong asymmetry in the induced magnetization is observed in the top versus bottom Pd layers (Figure \ref{fig:Kreuzpaintner_Figure_3}). The asymmetry is particularly large for very thin Pd capping layers, e.\,g.\ step 2, with the proximity induced magnetism being large and positive on the top surface, and much smaller on the bottom. With increasing thickness of the top Pd capping layer, approaching step 19, the magnetization profiles become increasingly symmetric. While an asymmetry in the magnetism is expected due to the island growth mechanism, the bottom surface is expected to possess the larger magnetization for all thicknesses due to its higher number density at the interface. Here, however, it possesses the smaller magnetization. Interestingly, the proximity induced magnetism extends 0.92\,nm into the Pd capping layer, and it is expected that a thickness larger than this will not influence the interfacial magnetism. However, with increasing Pd thickness beyond 0.92\,nm the asymmetry continues to decrease at the Pd/Fe/Pd interface. A symmetric magnetization on both sides of the Fe is only reached once the Pd capping layer thickness exceeds $\approx10$\,nm ($i \simeq 19$). This thickness also coincides with achieving a nearly identical nuclear density in each of the Pd layers, demonstrating the strong influence of structure on the magnetization profile.

The strong top/bottom asymmetry in the magnetization of the Pd, that we observed for the early growth stages of the Pd capping layer differs from the symmetric magnetization profile presented in \cite{doi:10.1103/PhysRevB.90.104403} where epitaxially grown samples were investigated. However, in \cite{doi:10.1103/PhysRevB.90.104403} the influence of the Fe layer thickness on the magnetization profile and ordering temperatures were in the focus of the study and the initial stages of Pd capping layer growth on the magnetization profile were not investigated. The top/bottom magnetic asymmetry, which is primarily located at the interface, and its dependence on the Pd thickness indicates the presence of both proximity-induced magnetism and implies Dzyaloshinskii-Moriya interactions \cite{doi:10.1103/PhysRevLett.115.267210,doi:10.1038/ncomms5030}. Specifically, an IDMI can be generated at HM/FM interfaces due to spin-orbit coupling and can be strong enough to induce the reorientation of the interfacial moments \cite{doi:10.1103/PhysRevB.93.235131}. The IDMI takes the form $E\propto(S_1 \times S_2)$, which motivates neighboring moments to orient at 90\,$^{\circ}$ with a well-defined handedness, due to the vector nature of the cross-product. As a result of the handedness, at the bottom Pd/Fe interface the IDMI will induce a curling of the magnetic moment in a direction determined by the polarity of the DMI coefficient. Although the DMI interaction at the interface of the 3d/4d transition metals is weak compared to the 3d/5d interfaces, it is not vanishingly small (-0.1 meV for Fe/Pd, compared to 1.7 meV for Fe/Ir) \cite{doi:10.1038/ncomms5030}. This curling will be balanced against the exchange energy – which prefers parallel alignment – and the DMI from the top surface of the Fe. Initially, the top surface of the Fe is vacuum, resulting in a vanishingly-weak value of the DMI. The asymmetry between the DMI at the top and bottom surfaces results in a curling of the magnetization, particularly at the Pd/Fe interface. Since the presented PNR results measure only the non-spin-flip channel, the measurements capture the projection of the magnetization along the direction of the guide field. The curling of the interfacial moments therefore manifest as a reduced magnetization compared to the bulk value. Subsequent deposition of a Pd capping layer – forming a Fe/Pd interface – will generate a curling of the magnetization in the opposite direction from the Pd/Fe interface due to the handedness of the DMI. The opposite curling from the top and bottom DMI cancel and allow the exchange term to dominate. Indeed, as the Pd capping layer thickness is increased, the magnetization at the Pd/Fe interface increases to agree with the expected value. For the thickest Pd capping layer, the DMI in the Pd/Fe/Pd structure is effectively symmetric, resulting in equal and opposite energies from the surfaces.

In detail, for our polycrystalline sample, after the Fe deposition step, the magnetization $M^{\text{Pd-1}}_{\text{i=A}}$ at the interface from Fe to the underlying Pd layer is comparatively small, showing a value of $M^{\text{Pd-1}}_{i=A} = 0.12$\,$\mu_{\text{B}}$/atom$^{\text{Pd}{\text{-1}}}$. Upon deposition of the Pd capping layer the induced magnetism in the Pd underlayer increases to $M^{\text{Pd-1}}_{i=19} \approx0.3$\,$\mu_{\text{B}}$/atom$^{\text{Pd}{\text{-1}}}$ at its final thickness. In comparison, the induced magnetism in the Pd capping layer, $M^{\text{Pd+1}}$, evolves opposite to $M^{\text{Pd-1}}$: with the first Pd deposition step, a magnetization of $M^{\text{Pd+1}}_{i=2} \approx0.6$\,$\mu_{\text{B}}$/atom$^{\text{Pd}{\text{+1}}}$ is observed, which gradually decreases with layer thickness to a final value of $M^{\text{Pd+1}}_{i=19} \approx0.3$\,$\mu_{\text{B}}$/atom$^{\text{Pd}{\text{+1}}}$.

Comparing these values with the literature, for the fully grown sample ($i=19$) the interfacial moment agrees well with previous works, which report an induced moment of $0.32~-~0.38$\,$\mu_{\text{B}}$/atom$^{\text{Pd}}$ for ideal interfaces \cite{doi:10.1103/PhysRevB.51.6364, doi:10.1007/BF00616980, doi:10.1016/S0304-8853(96)00480-5}. Interestingly, Ref.\ \cite{doi:10.1103/PhysRevB.51.6364} also reports a moment of $0.54$\,$\mu_{\text{B}}$/atom$^{\text{Pd}}$ for imperfect interfaces. This value agrees well with the initial deposition (e.\,g.\ step 2), in which we observe an induced moment of $\approx0.6$\,$\mu_{\text{B}}$/atom$^{\text{Pd}{\text{+1}}}$. This result is consistent with our expectation that the Pd film initially exhibits an island growth mode, effectively corresponding to a high surface roughness. Furthermore, a moment of $0.17$\,$\mu_{\text{B}}$/atom$^{\text{Pd}}$ \cite{doi:10.1007/BF00616980} and $0.26$\,$\mu_{\text{B}}$/atom$^{\text{Pd}}$ \cite{doi:10.1016/S0304-8853(96)00480-5} were reported for the second monolayer. These findings also agree well with the result of our \textit{in situ} study for the fully grown sample and if only the Pd capping layer is included in the investigations.

\section{Summary and Conclusions}
Using \textit{in situ} polarized neutron reflectometry, the evolution of the magnetism in a polycrystalline Pd(11\,nm)/Fe(0.41\,nm)/Pd(72\,nm) trilayer heterostructure was investigated during growth. An induced magnetization is observed in Pd up to 0.92\,nm from the Fe interface. It continuously decreases with increasing distance from the interface. Without any Pd capping layer (viz.\ Pd/Fe only), there is a small magnetic moment of $\approx0.12$ induced in the Pd underlayer ($\mu_{\text{B}}$/atom$^{\text{Pd-1}}$). Concurrently, the Fe magnetization is $1.6$\,$\mu_{\text{B}}$/atom$^{\text{Fe}}$, surprisingly small when compared with the bcc Fe bulk value of $2.2$\,$\mu_{\text{B}}$/atom$^{\text{Fe}}$ \cite{doi:10.1016/0375-9601(81)90311-X}. Subsequent deposition of the Pd capping layer further reduces the Fe moment to a final value of $\approx0.9$\,$\mu_{\text{B}}$/atom$^{\text{Fe}}$. This reduction is accompanied by an increase of the induced moment in the Pd underlayer at the interface, a decrease of the induced moment in the Pd capping layer, and an increase in the symmetry of the Pd/Fe/Pd magnetization profile. Interestingly, the symmetry of the magnetization profile is influenced by the Pd capping layer well beyond the interface, in regions which do not show induced magnetization themselves. Complementary TEM imaging of the Fe/Pd shows sharp interfaces, which indicates that the magnetization does not result from inter-diffusion of Fe and Pd at the interface.

Our observations as a whole indicate the presence of two effects in the investigated polycrystalline Pd(11\,nm)/Fe(0.41\,nm)/Pd(72\,nm) trilayer heterostructure, namely (i) proximity-induced magnetism extending into the Pd up to approximately 1\,nm on both sides of the Pd/Fe interface and (ii) a Dzyaloshinskii-Moriya interaction located directly at the Pd/Fe interface. The observations are similar as in the Co/Pt system, described in \cite{doi:10.1103/PhysRevLett.115.267210}, where no correlation between the presence of proximity-induced magnetism and Dzyaloshinskii-Moriya interaction was reported. However, as was shown here, the dominance of either effect and the magnetization for the Pd-1 and Fe layer, respectively, can experimentally be influenced by restoring the structural and electronic symmetry.

\section{acknowledgments}
This work is based on experiments performed at the Swiss spallation neutron source SINQ, Paul Scherrer Institute, Villigen, Switzerland. The funding of this project by Deutsche Forschungsgemeinschaft (DFG) within the Transregional Collaborative Research Center TRR\,80  ``From electronic correlations to functionality" is gratefully acknowledged. We thank Björgvin Hjörvarsson (Uppsala University) for fruitful discussions.
%\end{acknowledgments}

% Create the reference section using BibTeX:
\bibliography{Kreuzpaintner_bibliography}
\bibliographystyle{unsrt}
\end{document}